# Path of the current flow at the metal contacts of graphene field-effect transistors with distorted transfer characteristics

Ryo Nouchi[1,a)] and Katsumi Tanigaki[2,3]

[1]Nanoscience and Nanotechnology Research Center, Osaka Prefecture University, Sakai 599-8570, Japan

[2]WPI-Advanced Institute for Materials Research, Tohoku University, Sendai 980-8577, Japan

[3]Department of Physics, Graduate School of Science, Tohoku University, Sendai 980-8578, Japan

[a)] Electronic mail: r-nouchi@21c.osakafu-u.ac.jp

ABSTRACT: Graphene field-effect transistors with source/drain contacts made of metals that can be easily oxidized such as ferromagnetic metals often display a double dip structure in the transfer characteristics because of charge density depinning at the contacts. Generally, transfer characteristics of field-effect transistors show no dependence on the length of the source/drain contacts because charge carrier injection occurs mainly at the edges of the contact. However, the shape of the transfer characteristics of devices fabricated using Ni contacts is found to be dependent on the length of the contact. This peculiar behavior was attributed to charge carrier injection from near the center of the contacts. This is because of oxygen diffusion and the resultant formation of an interfacial oxide layer of non-uniform thickness. The observed contact length dependent transfer characteristics were reproduced using a model calculation that includes charge carrier injection from the center of the electrode and subsequent charge transport underneath the metal contact.

Graphene, which is a single layer of graphite, has a high charge carrier mobility and is regarded as a pivotal material for the development of high-speed electronics.[1] The conduction band minimum and the valence band maximum of graphene touch at the Dirac point where the electronic density of states is zero.[2] This unique electronic structure leads to ambipolar field-effect transistors (FETs) with transfer characteristics (the gate voltage dependence of the drain current) that typically display a V-shaped curve with a single dip. At the dip, the Fermi level of graphene coincides with the Dirac point. The positively- and negatively-gated branches with respect to the dip correspond to electron- and hole-conduction regions, respectively.

The shape of the transfer characteristics is affected by the source/drain metal contacts. The contacts are used to inject charge carriers into the graphene channel. The electrical conductance of the hole and electron branches becomes asymmetric because of charge transfer from the electrodes to the graphene.[3] Unlike the typical V-shaped curve, distorted transfer curves are observed in some cases because of an anomaly at the metal contacts.[4] The distortion appears as a double-dip structure, which can be understood as charge density depinning at the metal-graphene interface.[5,6] At normal metallic contacts with charge density pinning, the charge density near the contacts cannot be easily controlled with the applied gate voltage.[7] Depinning allows the modulation of the charge density in the graphene even at the metal contacts. Contact-induced doping of charge carriers occurs at the contacts,[8] and thus the Fermi level of the graphene at the contacts differs at positions far away from the contacts. The gate voltage that corresponds to the Dirac point in these two regions differs, leading to a transfer curve with a double-dip structure.[5,9]

The double-dip structure has been observed with metals that can be easily oxidized, including earth-abundant metals such as Al[10] and ferromagnetic metals such as Co[4] and Ni.[5,11] The latter metals may become critical for spintronic applications, where graphene is superior because of its long spin diffusion length.[12,13] The double-dip-type transfer characteristics in these systems can be attributed to the decoupling of direct metal-graphene interactions because of interfacial oxidation of the metal electrodes and the resultant charge density depinning.[5] The insertion of a tunnel barrier at the metal-graphene interface is an effective way to electrically inject a spin-polarized current from a ferromagnetic electrode into graphene.[14] The intentional insertion of an insulating tunnel barrier should also decouple direct metal-graphene interactions. Ferromagnetically-contacted graphene FETs with a thin tunnel barrier such as aluminum oxide[15] or MgO[16,17] were reported to display a double-dip structure in their transfer characteristics.

In this Letter, graphene FETs that display the double-dip-type transfer characteristics are investigated. The path that the current flows through the metal contacts with charge density depinning is determined by comparing the transfer characteristics of graphene FETs with different contact lengths. Here, the contact length is defined as the length of the source/drain electrodes along the channel length direction. Charge density depinning can be achieved by decoupling direct interactions between the metallic electrodes and the graphene underneath the electrodes. According to this mechanism, two types of metal contacts with charge density depinning are realized by intentionally inserting a molecular layer into the metal-graphene interface and the unintentional oxidation of the metal surfaces that come into contact with the graphene.[5] When a molecular layer was inserted, the path of current flow was similar to normal contacts with charge density pinning[18] and charge carrier injection occurred mainly from the edges of the electrode. However, the current flow through the oxidized metal contact occurs via charge carrier injection near the center of the electrodes. This is completely different from the near-edge injection postulated generally for various metal contact architectures.[19]

Flakes of single layer graphene were formed using mechanical exfoliation with adhesive tape onto a Si substrate with a 300 nm thick thermal oxide layer grown on top. The Si substrate was highly doped and was used as a back gate electrode. A single layer graphene flake was used which was determined using the optical contrast and Raman scattering spectra of the flakes. Electron-beam lithographic processes were used to pattern the source and drain electrodes on the graphene flakes. The electrodes were thermally evaporated in a high vacuum of the order of $10^{-4}$ Pa followed by a liftoff process. A 20 nm layer of Ni was used for the unintentional charge density depinning. Ni is a metal that is easily oxidized. A 50 nm thick Au layer with a 1 nm thick adhesion layer of Cr was used for the intentional depinning with an inserted molecular layer. The Au and Cr layers were deposited after the formation of the molecular layer. This combination of contacts (Au with a very thin Cr layer) was shown to display typical V-shaped transfer curves[4]. For the formation of the inserted molecular layer, a saturated solution of 7,7,8,8-tetracyanoquinodimethan (TCNQ) in toluene was dropped at room temperature onto the substrate following the lithographic patterning of an electron-beam resist. The solution was then blown by air 20 s after dropping. After the liftoff process, the substrates were rinsed using 2-propanol and acetone. Electrical characterization of the devices was conducted in air in the dark at room temperature.

Figure 1(a) shows the transfer characteristics of graphene FETs with Ni contacts with contact lengths of 0.8, 1.2 and 2.8 μm. All three transfer curves show a double-dip structure indicating the occurrence of charge density depinning. The sheet conductance defined by $I_D L/(W V_D)$ was plotted against the gate voltage, $V_G$, with respect to the Dirac point, $V_{NP}$. Here, $I_D$ is the drain current, $L$ is the channel length, $W$ is the channel width, and $V_D$ is the drain voltage. $V_D$ was set to 100 mV. The $V_{NP}$ values were set to the Dirac point of graphene in the channel. The Dirac point of the channel mainly reflects that of the channel center[20]. In the case of the Ni contact, electron donation from the contact is known to occur[5] and so the dip appeared at the right hand side that corresponded to the $V_{NP}$ of the channel. The left dip originating from regions with Ni contacts was more distinct as the contact length, $d$, becomes shorter.

Even if $d$ became shorter, the transfer characteristics should be independent of $d$ as long as charges are mainly injected from the edge of the electrode as shown schematically in Fig. 2(a).[19] An effective contact length for edge injection is known as the transfer length. As $d$ becomes shorter than the transfer length, the current injection turns into the areal injection, where the current flows through the whole metal-graphene interface. The transfer length of the metal-graphene contacts has been estimated to be ~1 μm[18] and so it would be expected that the shape of the transfer curves may change when $d$ becomes shorter than ~ 1 μm. However, these general considerations for metal contacts do not account for the results shown in Fig. 1(a) since the shape of the transfer curves of the FETs with Ni contacts changes when $d > 1$ μm.

To examine if this behavior is characteristic of charge density depinning, a different type of metal contact that displays the double-dip-type transfer characteristics was intentionally formed. This architecture ensures that the direct interactions between graphene and the electrode metal are decoupled using an inserted molecular layer. Figure 1(b) shows transfer characteristics of the TCNQ-inserted graphene FETs with contact lengths of 0.5, 1.0 and 2.0 μm. $V_D$ was set to 10 mV. TCNQ is an electron acceptor.[21] Holes are doped in the graphene and so a dip should appear in the positively-gated region. Although the dip corresponding to the Dirac point of the contacted region was not detected in the measured $V_G$ range, distortion was clearly seen in the positively-gated region. This indicates an occurrence of the charge density depinning. Unlike the devices with Ni contacts, the shapes of the transfer curves of the TCNQ-inserted device with $d$ of 1.0 and 2.0 μm are similar.[22] This feature is consistent with considerations for metal contacts in general. The behavior observed with the devices with Ni contacts cannot be explained using just charge density depinning.

A possible explanation for the behavior of the devices with Ni contacts is non-uniformity of the decoupling interlayer at the metal-graphene contact. In the case of intentional decoupling with the insertion of a TCNQ molecular layer, the layer was formed by simply dropping a solution of TCNQ molecules in toluene onto the graphene. The layer should be uniform in thickness on average, though microscopic inhomogeneity must exist because of the crystallization/aggregation of TCNQ molecules. In the case of the unintentional decoupling in the devices with Ni contacts, the interlayer is considered to be a metal oxide layer.[5] The electrode metal can be unintentionally oxidized by oxygen diffusion from the electrode edges. The thickness of the oxide layer is largest at the edges and gradually decreases in thickness towards the center. Thus, the least-resistance path should be through the thinnest point of the interfacial oxide layer. Current injection is considered to occur mainly from the center of the electrode as shown in Fig. 2(b). In this case, when $d$ becomes longer, charge carriers should travel a longer distance before entering the channel region. The distortion because of the charge density depinning becomes more distinct as shown in Fig. 1(a).

To verify this explanation, the transfer characteristics were calculated by mimicking the path of the current flow through the graphene FETs with Ni contacts as shown in Fig. 3(a). This calculation is based on an empirical model that was developed to characterize metal contact effects.[23] In short, the overall channel resistance, $R$, is expressed as a series connection of local resistances:

$$R = \frac{1}{W} \int_0^{L+d} \frac{1}{\sigma(x)} dx,$$

where $\sigma(x)$ is the local sheet conductance. The V shape of the $V_G$ dependence of $\sigma(x)$ can be mimicked by the function

$$\sigma(x) \equiv \sqrt{\{n(x)e\mu\}^2 + \sigma_{\min}^2} = \sqrt{\{\mu C_0 (V_G - V_{NP}(x))\}^2 + \sigma_{\min}^2},$$

where $n(x)$ is the local carrier concentration, $e$ is the elementary charge, $\mu$ is the carrier mobility, $\sigma_{\min}$ is the minimum sheet conductance that corresponds to the Dirac point, $C_0$ is the gate capacitance per unit area, and $V_{NP}(x)$ is the local Dirac point. The vertical axis in Fig. 3(b) is the sheet conductance, $G_{sq}$, given by $G_{sq} = (L + d)/(WR)$. It should be noted that $G_{sq}$ was calculated using the effective channel length, $L + d$, which includes carrier transport underneath the metal contacts. For simplicity, the doping level of the contacted region and the degree of charge density depinning[24] was assumed to be constant. The following parameters were used in the calculation: $L_D/L = 1/8$, $\mu$ was 1500 cm$^2$ V$^{-1}$ s$^{-1}$, the electron concentration because of doping at the contact was $4.0 \times 10^{12}$ cm$^{-2}$, the depinning factor[24] was 1 (complete depinning). $\sigma_{min}$ in the channel was set to $4e^2/h$ (ref. 25), where $h$ is Planck's constant. To take into consideration the contact-induced increase in the electronic density of states of the graphene,[26] $\sigma_{min}$ under the contact was set to $3 \times 4e^2/h$. The behavior observed experimentally with the devices with Ni contacts was reproduced as shown in Fig. 3(b). This supports carrier transport underneath the contacts.

In conclusion, the path that the charge carriers take at the source/drain metal contacts of graphene FETs with charge density depinning was investigated using transfer characteristics that were dependent on the contact length. In general, charge carriers are injected mainly from edges of metal contacts. This means that the shape of transfer curves is independent of the electrode contact length if the contact length is longer than the transfer length of the contact. Graphene FETs with electrodes that can be easily oxidized display a double-dip-type (or distorted) transfer curve because of charge density depinning at the metal-graphene contacts. The transfer curves of these devices were dependent on the contact length. This behavior can be explained if carriers were injected near the center of the contacts. The anomalous current flow could be explained by non-uniform oxidation of the electrode because of oxygen diffusion from the edges of the electrode to the center. This would lead to a non-uniform thickness of the interfacial oxide layer. The oxide layer would be thick at the edges and thin near the center. These results imply that oxygen can diffuse along the electrode-graphene interfaces for at least a few micrometers from the electrode edges. Oxygen intercalation has been reported in a graphene/Ru(0001) interface.[27] In this case the whole graphene Ru (001) interface with a lateral dimension of several tens of micrometers was fully oxidized after exposure to molecular oxygen ($10^{-7}$ Torr, 300 °C) for $10^3$ s in an ultrahigh-vacuum chamber.[27] The high diffusivity of oxygen along the metal contacts should be considered when fabricating graphene electronic devices with metal contacts since oxygen diffusion can drastically change electrical characteristics of the devices. A method to increase metal-graphene interactions or the use of a metallic species that adsorbs strongly onto graphene should be necessary to inhibit the oxygen diffusion.

This work was supported in part by the Special Coordination Funds for Promoting Science and Technology from the Ministry of Education, Culture, Sports, Science and Technology of Japan and by a technology research grant from JFE 21st Century Foundation. The graphite crystal used in the present study was supplied by M. Murakami and M. Shiraishi.

**Figures**

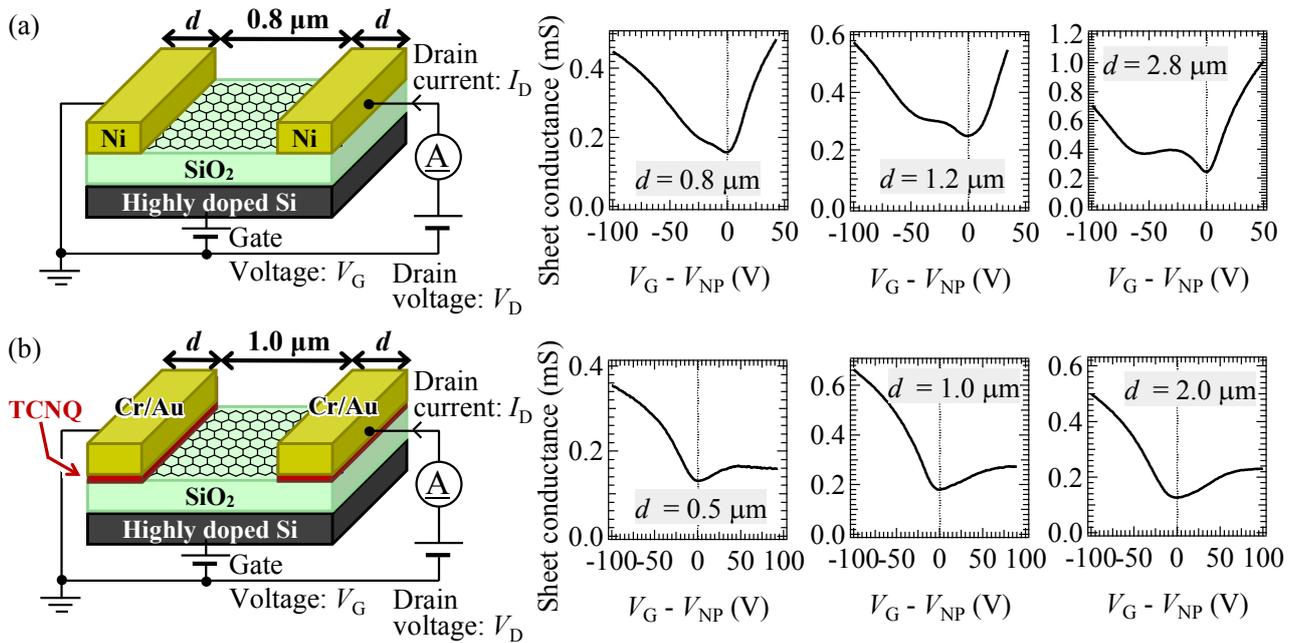

FIG. 1. A schematic of the fabricated graphene FETs and the measured transfer characteristics of FETs with (a) Ni contacts and (b) metal contacts with the insertion of TCNQ. Devices with various contact lengths were investigated.

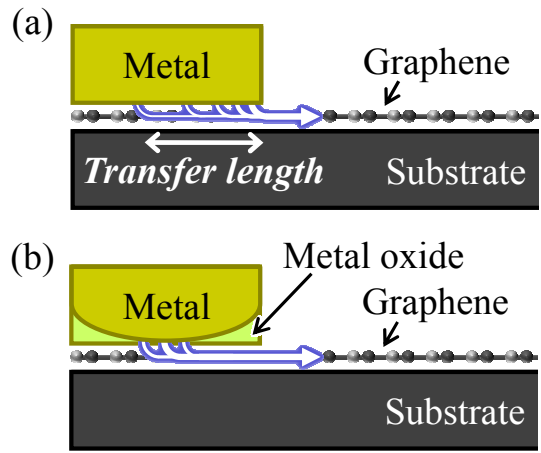

FIG. 2. The flow of current through the metal-graphene interface for (a) standard metal contacts and (b) metal contacts that can be easily oxidized.

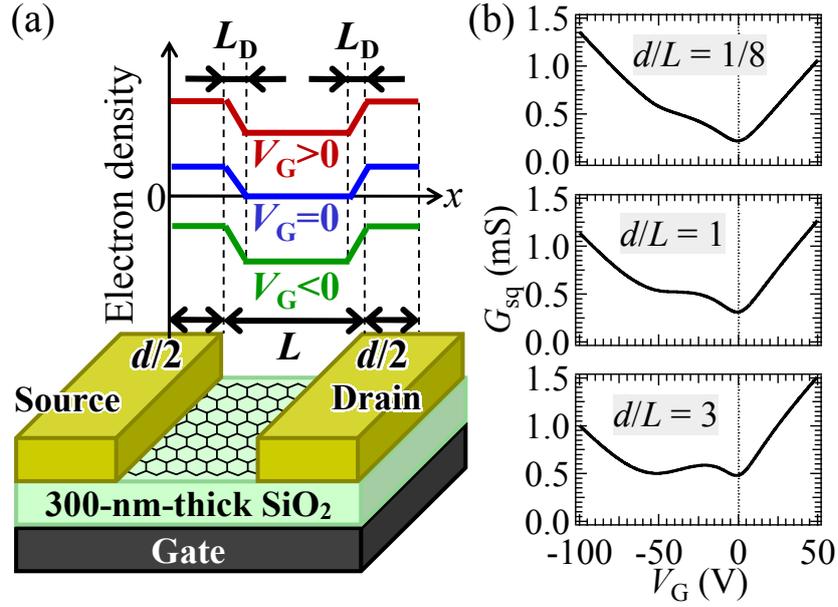

FIG. 3. The model calculation of the transfer characteristics of graphene FETs with charge density depinning and carrier transport underneath metal contacts. (a) The assumed carrier density profile along the carrier transport path and its gate-voltage dependence. Complete depinning at the contacts is assumed. (b) The calculated transfer characteristics for various contact lengths. In this calculation, $L_D/L$ is 1/8, the electron and hole mobilities are 1500 cm$^2$ V$^{-1}$ s$^{-1}$ and the electron concentration because of doping at the metal contact was $4.0 \times 10^{12}$ cm$^{-2}$. The minimum sheet conductance of graphene in the channel and under the contact was set to $4e^2/h$ and $12e^2/h$, respectively.